\documentclass[final,5p,times,twocolumn]{elsarticle}

 \usepackage{epsfig}

\usepackage{amssymb}

\journal{Physics Letters B}

\begin{document}

\begin{frontmatter}



\title{Anti-analog giant dipole resonances and the neutron skin of nuclei}


\author[atomki]{A. Krasznahorkay}
\ead{kraszna@atomki.hu}
\author[uz]{N. Paar}
\author[uz]{D. Vretenar}
\author[kvi]{M.N. Harakeh}

\address[atomki]{Institute for Nuclear Research, Hungarian Academy of Sciences
  (MTA Atomki), H-4001 Debrecen, P.O. Box 51,
  Hungary}
\address[uz]{Physics Department, Faculty of Science, University of Zagreb,
  Croatia}
\address[kvi]{Kernfysisch Versneller Instituut, University of Groningen,
  Groningen, The Netherlands}

\begin{abstract}
We examine a method to determine the neutron-skin thickness of 
nuclei using data on the charge-exchange anti-analog giant dipole resonance (AGDR).
Calculations performed using the relativistic proton-neutron quasiparticle 
random-phase approximation (pn-RQRPA) reproduce the isotopic trend of the excitation
energies of the AGDR, as well as that of the spin-flip giant dipole resonances (IVSGDR), in 
comparison to available data for the even-even isotopes $^{112-124}$Sn.
It is shown that the excitation energies of the AGDR, obtained using  a set of 
density-dependent effective interactions which span a range 
of the symmetry energy at saturation density, supplemented with 
the experimental values, provide a stringent constraint on value of the neutron-skin thickness.
For  $^{124}$Sn, in particular, we determine the value $\Delta R_{pn}$=0.21$\pm$0.05 fm.
The result of the present study shows that a measurement of the excitation energy of
the AGDR in ($p$,$n$) reactions using rare-isotope beams in inverse kinematics, 
provides a valuable method for the determination of
neutron-skin thickness in exotic nuclei.

\end{abstract}


\end{frontmatter}




\section{Introduction}

An interesting phenomenon in nuclear structure is the formation 
of a skin of neutrons on the surface of a nucleus, and its 
evolution with mass number in an isotopic chain \cite{su95}.
A precise measurement of the thickness of neutron skin is important 
not only because this quantity represents a basic nuclear property, 
but also because its value constrains the symmetry energy term of the
nuclear equation of state \cite{te08,ta11,ro11,ab12,VNPM.12,Pie12}.
A detailed knowledge of the symmetry energy is essential for describing
the structure of neutron-rich nuclei, and for modeling properties of neutron-rich
matter in applications relevant for nuclear astrophysics.

The difference between the neutron and proton rms radii is rather
small (few percent) and a precise measurement of the neutron-skin thickness
presents a considerable challenge.
Several  methods have been used to determine this quantity \cite{te08,ba89,kr91,kr04,tr01,br07,ts12}, 
but almost all of these are applicable only to stable nuclei and the results are model dependent \cite{kr04}.
Methods based on coherent nuclear motion include excitations of the isovector giant dipole 
resonance (IVGDR) \cite{kr91}, the isovector spin giant dipole resonance (IVSGDR) \cite{kr04}, 
the Gamow-Teller resonance (GTR) measured relative to the Isobaric Analog State (IAS) \cite{Vre03}, 
and high-resolution study of the electric dipole polarizability \cite{ta11}.

The Pb Radius Experiment (PREX) at JLAB \cite{ab12} has initiated a new line of 
research based on the parity-violating elastic electron scattering to measure the
neutron density radius $R_n$, which in turn allows to determine the neutron-skin
thickness from $\Delta R_{pn} = R_n - R_p$, where $R_p$ is the radius of the proton
density distribution. 
Although parity-violating elastic electron scattering provides a model independent 
measurement of $\Delta R_{pn}$, its current precision is far from satisfactory and the 
method cannot be applied to unstable isotopes.

Radioactive ion beams (RIBs) have recently been employed 
to determine the neutron-skin thickness in unstable nuclei, 
specifically in measurements
of reaction cross sections and pygmy dipole 
resonances \cite{su95,kl07,ca10}.
For an accurate determination of this quantity using RIBs, 
it is imperative to find a feasible method that employs 
reactions with low-intensity RIBs in inverse kinematics.
We have recently introduced a new method 
\cite{kr12,kr12a}
based on the excitation of the anti-analog giant dipole
resonance (AGDR)
observed in ($p$,$n$) reaction \cite{st80}. As pointed out by Krmpoti\' c, the excitation 
of the AGDR depends
sensitively on the neutron-skin thickness \cite{kr83} and, therefore,
$\Delta R_{pn}$ could be deduced from the measurement
of AGDR excitation energy. 

The main objective of this work is to test the method that determines the neutron-skin thickness in 
nuclei from AGDR data. By calculating excitation energies E(AGDR) and $\Delta R_{pn}$
in a fully self-consistent theoretical approach, and comparing to available
data, the feasibility of the method will be tested with the aim to
provide a basis for future studies with RIBs.

The paper is organized as follows.
In Sec. 2 we present a short outline of the AGDR mode and discuss its properties with the aim
to constrain the neutron-skin thickness. The experimental results on AGDR and IVSGDR 
obtained from ($p$,$n$) and ($^3$He,$t$) reactions for Sn isotopes are reviewed in Sec. 3. The 
theoretical framework and model calculations of 
excitation energies of  the AGDR and IVSGDR, and neutron-skin thickness for the even
Sn-isotopes, are described in Sec. 4.  The centroid of the AGDR for $^{124}$Sn is
deduced from an earlier ($p$,$n$) measurement in Sec. 5, and used to 
determine the corresponding neutron-skin thickness by comparing with model calculations.
Section 6 includes the summary and a brief outlook for future studies.

%
%
\section{Charge-exchange AGDR and neutron-skin thickness}

\begin{figure}[htb]
\includegraphics[width=80mm]{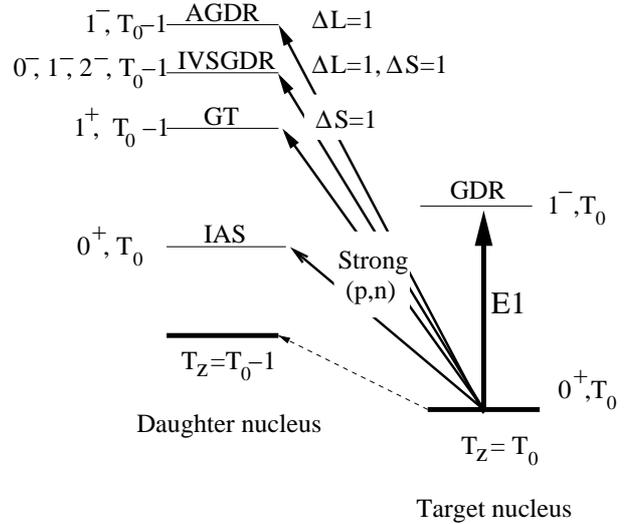}
\caption{The ground state and the GDR of the target nucleus ($T_z =
  T_0$). Also shown are the IAS (isospin=$T_0$) and anti-analog states
  (isospin=$T_0-1$) GTR, IVSGDR and AGDR in the daughter nucleus ($T_z =
  T_0-1$), excited in a ($p$,$n$) reaction. Only these components are shown
  because they are strongly favored by isospin selection rules for large
  $T_0$. See also Ref. \cite{st80}.
\label{levels}}
\end{figure}

The AGDR corresponds to the $\Delta J^{\pi}=1^-$, $\Delta L = 1$ resonant excitation,
and represents the anti-analog giant dipole resonance, i.e., the $T_0 - 1$ component of the
charge-exchange GDR ($T_0$ is the ground-state isospin of the
target nucleus). Figure \ref{levels} illustrates the ground state 
and the giant dipole resonance (GDR) state of a target nucleus ($T_z=T_0$), 
and the corresponding resonant states in the daughter nucleus reached by the
($p$,$n$) charge-exchange reaction: the IAS (isospin = $T_0$),  and the anti-analog (isospin = $T_0 - 1$)
states: GTR, IVSGDR, and AGDR \cite{st80}. 

The transition strength of dipole excitations
 is fragmented into the $T_0 - 1$, $T_0$ and
$T_0 +1$ components because of the isovector nature of the ($p$,$n$)
reaction.  The
pertinent Clebsch-Gordan coefficients \cite{os92} show that the $T_0 -1$
component (AGDR) is favored with respect to the $T_0$ and $T_0+1$ components,
by factors of $T_0$ and $2T_0^2$, respectively. Accordingly, Fig. \ref{levels} shows only
components that are strongly favored by isospin selection rules for large $T_0$ \cite{st80}.

For the charge-exchange dipole operator
\begin{equation}
\hat{O}_{\pm} = \sum_i r_i Y_{10}(\hat{r}_i)\tau^{(i)}_{\pm},
\label{operator}
\end{equation}
where $\tau^{(i)}_{\pm}$ denotes isospin raising and lowering operators, one obtains
the non-energy-weighted sum rule (NEWSR) \cite{au81},
\begin{equation}
S^- - S^+ =\frac{1}{2 \pi} (N\langle r_n^2\rangle - Z\langle r_p^2\rangle).
\label{srule}
\end{equation}
$S^-$ and $S^+$ denote the sums of transition strengths in the $\beta^-$ and
$\beta^+$ channels calculated using Eq.~(\ref{operator}), respectively. The AGDR mode is mediated by
the $\hat{O}_{-}$ operator. We note that the sum rule Eq.~(\ref{srule}) is proportional to the one
of the IVSGDR 
(up to the factor three that corresponds to different spin components in the latter case), previously
used to determine the neutron-skin thickness \cite{kr04,kr99}.
%
%
%
In Ref.~\cite{au81}, Auerbach et al. derived an energy-weighted sum 
rule (EWSR) for the dipole strength excited in charge-exchange reactions. 
However, this EWSR is not sensitive to the neutron-skin thickness
\cite{au81}.

Although the NEWSR provides information on the neutron-skin thickness, in practice it is not straightforward to determine 
$\Delta R_{pn}$ from the sum rule. 
The problem is that usually the non-energy weighted strengths are not simultaneously
available for both $\beta^-$ and $\beta^+$ channels, i.e., experiments
are mainly performed in the $\beta^-$ channel and provide $S^-$ only. 
To deduce $\Delta R_{pn}$ from the NEWSR additional approximations or
theoretical input are needed, e.g. the estimate for $S^+$ and the
normalization constant \cite{kr04}. One can, however, intuitively understand
the relation between the neutron-skin thickness and the energy of the AGDR if
one considers that for nuclei with $N-Z \gg 1$ one can neglect $S^+$ because
of Pauli blocking. In such a case, the EWSR, which is a constant, can be
expressed as the product of the NEWSR and the AGDR energy and thus one can
easily understand the inverse proportionality between the AGDR energy and
neutron-skin thickness, i.e. $E_{AGDR}$ decreases if $\Delta R_{pn}$ increases
and vice versa.

Instead of the NEWSR constraint on $\Delta R_{pn}$, in this work 
we aim to establish an alternative approach motivated by 
the study of Ref.~ \cite{kr83}, where a simple schematic model indicated 
strong sensitivity of the AGDR excitation energy on $\Delta R_{pn}$. In Sec. 4
we will explore this relation using a fully microscopic theoretical approach and 
available data on the AGDR energies.

\section{Isovector giant resonances excited by ($p$,$n$) reactions}

The first identification of a giant dipole transition excited in charge-exchange ($p$,$n$)
reactions was reported by Bainum et al. \cite{ba80} for the case $^{90}$Zr($p$,$n$)$^{90}$Nb 
at 120 MeV. In addition to the pronounced excitation of the GTR, a
broad peak was observed at
an excitation energy of 9 MeV above the GTR, with an angular distribution
characteristic of a $\Delta L=1$ transfer. This excitation energy is about 4
MeV below the location of the $T=5$ analog of the known GDR in $^{90}$Zr, and
thus it was suggested that this state is the $T=4$ anti-analog of the GDR.

Dipole resonances have also been studied systematically in ($p$,$n$) reactions at E$_p =
45$ MeV by Sterrenburg et al. \cite{st80}, for 17 different targets from
$^{92}$Zr to $^{208}$Pb.  Nishihara et al. \cite{ni85} measured also the
dipole strength distributions at E$_p = 41$ MeV.  It was shown
\cite{os81,au01} that the observed $\Delta L$= 1 resonance in general
corresponds to a superposition of all possible 
spin-flip dipole (IVSGDR)
and non-spin-flip dipole (AGDR) modes.
According to Osterfeld \cite{os92}, the
non-spin-flip to spin-flip ratio is favored at low bombarding energy (below 50
MeV), and also at very high bombarding energy (above 600 MeV).  Properties of
the IVSGDR were further investigated by Gaarde et al.  \cite{ga81} using
($p$,$n$) reactions on targets with mass of $40 \leq A \leq 208$, and by Pham et
al. \cite{ph95} in ($^3$He,$t$) reactions. In every experimental spectrum a
peak was observed at an energy several MeV above the GTR, with an angular 
distribution characteristic of a $\Delta L=1$ transfer.  The observed
excitation energies of the AGDR \cite{st80} and IVSGDR \cite{kr99,ph95},
relative to the IAS, are shown in Fig. 2 as functions of the mass number for
the even-even Sn isotopes.
For both modes one observes a systematic decrease 
of the excitation energy along the Sn isotopic chain.

\begin{figure}[htb]
\includegraphics[width=90mm]{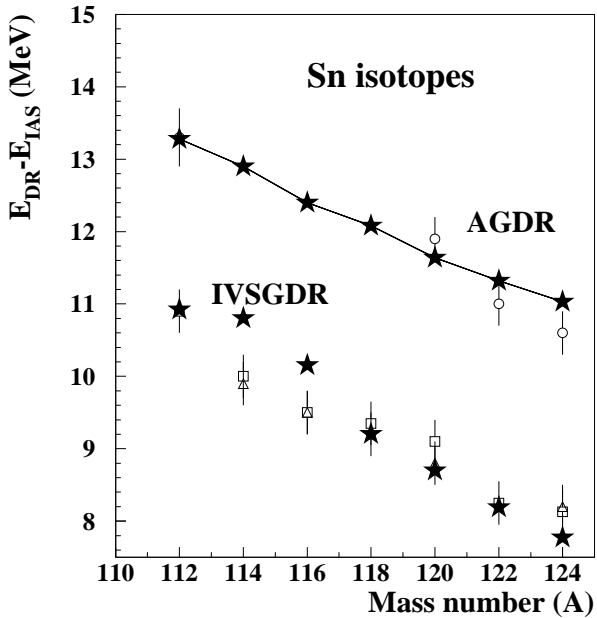}
\caption{Excitation energies of the AGDR and IVSGDR relative to the IAS 
for the even-even Sn isotopes as
  functions of the mass number.  The circles represent the experimental
  results from Sterrenburg et al. \cite{st80}, the squares are from Pham et
  al. \cite{ph95}, the triangles from Krasznahorkay et al. \cite{kr99}, and
  the stars are the pn-RQRPA values for the AGDR (higher) and
  the IVSGDR (lower), calculated with the DD-ME2 effective
  interaction~\cite{LNVR.05}.
\label{dr-ene}}
\end{figure}
 
The ($p$,$n$)  reaction has a high cross-section, and in inverse kinematics
the energy of the neutrons is only a few MeV, which can be measured with
highly efficient detectors. In our recent experiments \cite{kr12}, a
600 MeV/nucleon $^{124}$Sn relativistic heavy-ion
beam was directed onto a hydrogen target. 
The ejected neutrons were detected by a low-energy neutron-array (LENA)
ToF spectrometer \cite{la11,st13}, developed in Debrecen. The spectrometer was
placed at 1 m from the target, covering a laboratory scattering-angle region
of 65$^\circ\leq\Theta_{LAB}\leq75^{\circ}$. 
In this way, the excitation energy of the AGDR and IAS can be
determined, and new data enable studies
of neutron-skin thickness in nuclei. One expects that future progress with RIBs and
novel experimental techniques should provide data on the AGDR in exotic 
nuclei with even more pronounced neutron skin.
The experimental feasibility of the suggested method is also supported by a
recent publication in which the strength distribution of the Gamow-Teller giant
resonance was studied by the ($p$,$n$) reaction with RIBs
\cite{sa11} using a similar neutron spectrometer \cite{pe12}.

\section{Theoretical analysis}
To describe the evolution of excitation energies of the AGDR relative to 
the IAS, and their relation to $\Delta R_{pn}$, we perform a 
microscopic theoretical analysis based on relativistic nuclear 
energy density functionals.
The theoretical framework is realized in terms of the 
fully self-consistent
relativistic proton-neutron quasiparticle random-phase approximation
(pn-RQRPA) based on the relativistic Hartree-Bogoliubov model (RHB)
\cite{VALR.05}.  The pn-RQRPA is formulated in the canonical single-nucleon basis
of the RHB model in Ref.~\cite{Paar2003}, and extended to the description of
charge-exchange excitations (pn-RQRPA) in Ref.~\cite{Paar2004}. The RHB +
pn-RQRPA model is fully self-consistent: in the particle-hole channel
effective Lagrangians with density-dependent meson-nucleon couplings are
employed, and pairing correlations are described by the pairing part of the
finite-range Gogny interaction~\cite{BGG.91}.


For the purpose of the present study we employ a family of density-dependent
meson-exchange (DD-ME) effective interactions for which the constraint on the
symmetry energy at saturation density was systematically varied, and the
remaining model parameters were adjusted to reproduce empirical nuclear-matter
properties (binding energy, saturation density, compression modulus), and the
binding energies and charge radii of a standard set of spherical nuclei
\cite{VNR.03}.  These effective interactions were used to provide a
microscopic estimate of the nuclear-matter incompressibility and symmetry
energy in relativistic mean-field models~\cite{VNR.03}, and in
Ref.~\cite{kl07} to study a possible correlation between the observed pygmy
dipole strength in $^{130,132}$Sn and the corresponding values for the
neutron-skin thickness. In addition to a set of effective interactions with
systematically varied values of the symmetry energy at saturation density, the
relativistic functional DD-ME2 \cite{LNVR.05} is also used here to calculate
the excitation energies of the AGDR with respect to the IAS, as a function of
the neutron skin.  Pertinent to the present analysis is the fact that the
relativistic RPA with the DD-ME2 effective interaction predicts for the dipole
polarizability
\cite{ta11}
\begin{equation}
\alpha_D = {{8 \pi}\over 9} e^2~m_{-1}
\label{dip-pol}
\end{equation}
(directly proportional to the inverse energy-weighted moment $m_{-1}$)
of $^{208}$Pb the value $\alpha_D$=20.8 fm$^3$, in very good
agreement with the recently measured value: $\alpha_D = (20.1\pm 0.6)$
fm$^3$ \cite{ta11}.

In addition to the experimental excitation energies, Fig. 2 also includes the 
theoretical results obtained with the RHB + pn-RQRPA
model using the DD-ME2 effective interaction. The difference in the excitation
energy of the AGDR and the IAS, as well as between the IVSGDR and the IAS, for
the even-even Sn isotopes are shown as functions of the mass number.  For the
excitation energies of the AGDR and the IVSGDR we take the centroids of the
theoretical strength distributions: $m_1/m_0$, whereas a single peak is
calculated for the IAS.  Within the experimental uncertainty, we find 
a reasonable agreement between the data and the theoretical values for the AGDR.
The largest deviations, $\approx 0.4$ MeV, correspond to $^{122,124}$Sn. 
The agreement is less satisfactory for the IVSGDR, with the
discrepancy being especially large for $^{114}$Sn and $^{116}$Sn.  Part of the
difference is probably caused by the overlapping of the two resonances.
Experimentally it is possible to enhance the excitation of the AGDR with
respect to the IVSGDR by choosing a low bombarding energy ($\leq$ 45 MeV), but
it is expected that the suppressed IVSGDR can still cause some lowering of the
energy centroid of the AGDR observed in the ($p$,$n$) reaction since the former
has a lower excitation energy.  On the other hand, the excitation of the
IVSGDR can be enhanced at higher bombarding energy (around 150 - 200 MeV) but
a small fraction of the AGDR still remains, raising the centroid energy as the
energy of the AGDR is higher than that of the IVSGDR.  Since the IVSGDR has
three components with $\Delta J^\pi = 0^-, 1^-$ and $ 2^-$, its strength generally
spreads over a larger energy interval compared to the AGDR. The model
calculation cannot reproduce these structures as precisely as the centroid
energy of the AGDR, and this is the reason why in this work we make use of the
AGDR to determine the neutron-skin thickness.

The calculated values of the neutron-skin thickness for the Sn isotopes as a
function of mass number are compared to available data in Fig. 3. The RHB
neutron-skin thicknesses obtained using the DD-ME2 effective interaction are in very good
agreement with the experimental values obtained using various methods
\cite{te08,kr04,tr01}.
The self-consistent RHB calculation of $\Delta R_{pn}$,
and the corresponding pn-RQRPA excitation energies of the AGDR, 
establish a connection between these quantities and suggest 
a feasible method for determining the neutron-skin
thickness from AGDR data.

 \begin{figure}[htb]
\includegraphics[width=90mm]{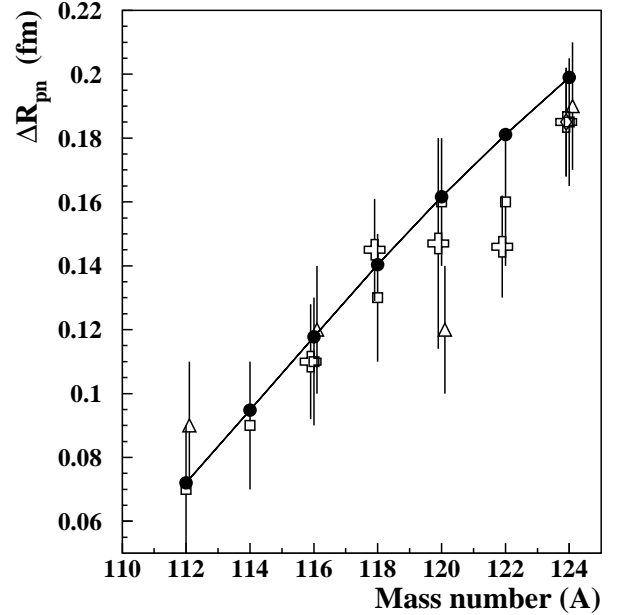}
\caption{Calculated values of the neutron-skin thickness for the even-even Sn
  isotopes as a function of the mass number (filled circles connected by the solid
  line), compared to experimental results obtained with the antiproton
  absorption method \cite{tr01} (triangles), from ($p$,$p$)
  scattering data \cite{te08} (crosses), and with the IVSGDR method \cite{kr04}
  normalized to the ($p$,$p$) result for $^{124}$Sn \cite{te08} (squares).
\label{skin}}
\end{figure}

\section{Determination of the neutron-skin thickness of $^{124}$Sn}

In this section the measured AGDR excitation energy
for $^{124}$Sn, together with 
the consistent RHB plus pn-RQRPA model calculation 
of $\Delta R_{pn}$ and the AGDR energy, is used 
to constrain the value of the neutron-skin thickness.
We consider the available
data for the AGDR for $^{124}$Sn from Sterrenburg et al. \cite{st80} ($E(AGDR)
- E(IAS) = 10.60 \pm 0.20 $ MeV), but slightly increased to $E(AGDR) - E(IAS)
= 10.93 \pm 0.20 $ MeV in order to approximately compensate the effect of the
energy shift caused by the mixing with the IVSGDR. Below we explain how this
energy shift is determined.

Austin et al. \cite{au01} developed a phenomenological model to describe the
variation with bombarding energy of the peak positions of the AGDR and IVSGDR
observed in ($p$,$n$) reactions.  They assumed that the position C of the
centroid of the
$\Delta L=1$
excitations (including both the AGDR and IVSGDR) at a
bombarding energy $E_p$ is given by the weighted average of the energies:
\begin{equation}
C={\sigma_0E_0 + \sigma_1E_1\over\sigma_0 + \sigma_1} =E_0 -
{\sigma_1/\sigma_0\over 1 + \sigma_1/\sigma_0}\Delta \ ,
\end{equation}
\noindent where $E_0 (E_1) $ is the energy of the AGDR (IVSGDR),
$\Delta=E_0 - E_1$ and $\sigma_0 (\sigma_1)$ is the cross
section for $S=0$ $(S=1)$ transfer.  They estimated the $\sigma_1/\sigma_0$ ratio
by : $\sigma_1/\sigma_0 \approx (E_p(MeV)/55)^2$ \cite{au01} and obtained the
energy of the AGDR in $^{124}$Sn to be 14.4 $\pm$ 2.2 MeV, which is completely
different from any theoretical prediction \cite{au01}.  In reality, the
centroid of the dipole strength distribution is usually determined by fitting
the distribution by a Gaussian or a Lorentzian curve.  This makes a
significant difference in case of $^{124}$Sn, where the AGDR and the IVSGDR
display very different widths: 3.6 MeV \cite{st80} and 9 MeV \cite{ph95},
respectively.

To determine the energy shift of the AGDR peak at $E_p$ = 45 MeV from the
empirical peak energy, we simulate the mixing of the AGDR and IVSGDR by using
their real widths of 3.6 MeV and 9 MeV, the ratio of their intensities as
approximated by Austin et al. \cite{au01}, and their energy difference
$\Delta$=2.3 MeV obtained from Fig. 2. The composite spectrum is then fitted
by a Gaussian curve in a reasonably wide energy range ($\pm$ 5 MeV) around the
position of the peak, and this yields an energy shift of 0.33 MeV for the
AGDR.

The sensitivity of the centroid energy of the AGDR to the neutron-skin
thickness of $^{124}$Sn is explored by performing RHB + pn-RQRPA calculations
using a set of the effective interactions with different values of the
symmetry energy at saturation: $a_4 =$ 30, 32, 34, 36 and 38 MeV (and
correspondingly different slopes of the symmetry energy \cite{VNPM.12}) and,
in addition, the DD-ME2 effective interaction ($a_4 =32.3$ MeV).  In Fig. 4,
the resulting energy differences $E(AGDR) - E(IAS)$ are plotted as a function
of the corresponding neutron-skin thickness $\Delta R_{pn}$ predicted by these
effective interactions.

\begin{figure}[ht]
\includegraphics[width=80mm]{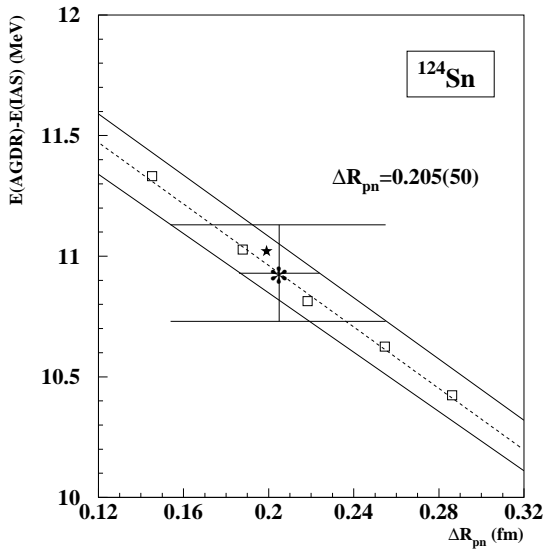}
\caption{The difference 
in the excitation energy of the AGDR and the IAS for the 
target nucleus $^{124}$Sn, calculated with the pn-RQRPA 
using five relativistic effective interactions characterized 
by the symmetry energy at saturation 
$a_4 =$ 30, 32, 34, 36 and 38 MeV (squares), and the interaction 
DD-ME2 ($a_4 =32.3$ MeV) (star). The theoretical values $E(AGDR) - E(IAS)$
are plotted as a function of the corresponding ground-state neutron-skin 
thickness $\Delta R_{pn}$, and compared to the experimental 
value.  
\label{skin124}}
\end{figure}

The two parallel solid lines in Fig. 4 delineate the region of theoretical
uncertainty for the used set of effective interactions. When adjusting the
parameters of these interactions \cite{LNVR.05,VNR.03}, an uncertainty of 10\%
was assumed for the difference between neutron and proton radii for the nuclei
$^{116}$Sn, $^{124}$Sn, and $^{208}$Pb.  This set of interactions was also
used to calculate the electric dipole polarizability and neutron-skin
thickness of $^{208}$Pb, $^{132}$Sn and $^{48}$Ca, in comparison with the
predictions of more than 40 non-relativistic and relativistic mean-field
effective interactions \cite{Pie12}.  

By comparing the experimental result for $E(AGDR) - E(IAS)$ to the theoretical
energy differences (see Fig. 4), we deduce the value of the neutron-skin
thickness in $^{124}$Sn: $\Delta R_{np} = 0.205 \pm 0.050 $ fm (including
theoretical uncertainties).  In Table I the value for $\Delta R_{np}$
determined in the present analysis is compared to previous results obtained
with a variety of experimental methods.  The very good agreement with
previously determined values reinforces the expected reliability of the
proposed method.

\begin{table}[h]
\caption{\label{tab:table1} Values of the neutron-skin thickness 
($\Delta R_{pn}$) of
  $^{124}$Sn determined using various experimental methods, in comparison with 
  the neutron-skin thickness deduced in the present work. }
\begin{tabular}{lllc}
\hline\hline
\textrm{Method}& \textrm{Ref.}& \textrm{Date}& \textrm{$\Delta R_{pn}$} (fm)\\ 
\hline
 ($p$,$p$) 0.8 GeV & \cite{ba89} & 1979 & 0.25 $\pm$ 0.05 \\ 
($\alpha, \alpha$') IVGDR 120 MeV & \cite{kr94} & 1994 & 0.21 $\pm$ 0.11\\ 
antiproton absorption & \cite{tr01,br07} & 2001 & 0.19 $\pm$ 0.09 \\
($^3$He,$t$) IVSGDR+AGDR & \cite{kr04} & 2004 & 0.27 $\pm$ 0.07\\ 
pygmy dipole resonance         & \cite{kl07,ts12} & 2007 & 0.19 $\pm$ 0.05 \\
($p$,$p$) 295 MeV & \cite{te08,ts12} & 2008 & 0.185 $\pm$ 0.05 \\ 
AGDR \ present result & & 2013 &  0.21  $\pm$ 0.05 \\
\hline
\end{tabular}
\end{table}

\section{Conclusion and outlook} 
 A method to determine the size of the neutron-skin thickness in 
nuclei using data on the anti-analog giant dipole resonance has been discussed.
Charge-exchange $(p$,$n)$ reactions provide an excellent probe for the
neutron-skin thickness, as already demonstrated by measurement of the
IVSGDR and GTR, and the AGDR provides a complementary approach. 
In contrast to the IVSGDR, which displays a complex underlying 
structure with three overlapping components and its strength spreads over a large 
energy interval, the AGDR represents a rather simple charge-exchange
mode ($\Delta J^{\pi}=1^-$, $\Delta L = 1$). While previous analyses 
were based on sum rules, this work introduces an alternative self-consistent
approach that could systematically be used not only for the AGDR, but also
for other modes sensitive to the neutron skin.

As a first test, we have used the self-consistent RHB plus proton-neutron RQRPA to
calculate the $T_0-1$ dipole excitations in a sequence of Sn isotopes.  By
using effective interactions with density-dependent meson-nucleon couplings
in the particle-hole channel, and the pairing part of the Gogny interaction
D1S for the T=1 pairing channel, it has been possible to reproduce the
experimental results on the excitation energy of the AGDR relative to the
isobaric analog state. We have also shown that the isotopic dependence of the
energy difference between the AGDR and IAS provides direct information on the
evolution of neutron-skin thickness along the Sn isotopic chain. Very good
results have been obtained in comparison with available data on the
neutron-skin thickness. The present analysis demonstrates that this quantity
can be determined by measuring the excitation energies of the AGDR relative to
IAS.

The accuracy of the method has been tested in the example of $^{124}$Sn.  By
employing a set of effective interactions that span a broad range of values
for the neutron-skin thickness (as a result of variation of the symmetry
energy at saturation density), the size of the neutron skin has been
determined from the AGDR energies relative to IAS. The result 
is in very good agreement with previously published experimental
values. More extensive studies, in line with recent work on the electric dipole polarizability
and neutron-skin thickness \cite{Pie12} that has employed families of non-relativistic
and relativistic energy density functionals, would allow a further reduction of theoretical
uncertainties.
The successful test of the method based on the AGDR holds promise  
for determining the size of the neutron-skin of unstable 
neutron-rich exotic nuclei, and
this is reinforced by recent advances in the development of RIBs.

\section{Acknowledgement}  This work has been supported by the European
Community FP7 - Capacities, contract ENSAR n$^\circ$ 262010, the Hungarian OTKA
Foundation No.\, K106035, by the MZOS - project 1191005-1010, and the Croatian 
Science Foundation.

\section{References}

\bibliographystyle{elsarticle-num}
\bibliography{<your-bib-database>}

\end{document}